\documentclass[aps,prd,twocolumn,amsmath,amssymb,floatfix,nofootinbib,superscriptaddress]{revtex4-1}
\usepackage{graphicx}
\usepackage{dcolumn}
\usepackage{bm}
\usepackage[usenames, dvipsnames]{color}
\usepackage[normalem]{ulem}

\usepackage[colorlinks,bookmarks]{hyperref}
\definecolor{linkblue}{rgb}{0,0,0.8}
\definecolor{linkgreen}{rgb}{0,0.5,0}

\hypersetup{pdfpagemode=UseNone, pdfstartview=FitH, linkcolor=linkblue, %
            citecolor=linkgreen, urlcolor=linkblue}

\def\la{\langle}
\def\ra{\rangle}
\def\beq{\begin{equation}}
\def\eeq{\end{equation}}
\def\d{\partial}

\newcommand{\vl}{\boldsymbol{\ell}}
\newcommand{\vn}{\boldsymbol{n}}
\newcommand{\hvn}{\hat{\vn}}

\newcommand{\vL}{\boldsymbol{L}}
\newcommand{\vtheta}{\boldsymbol{\theta}}

\newcommand{\lmax}{\ell_{\rm max}}

\newcommand{\nhat}{\hat{ \mathbf{n}}}

\def\fnllocal{f_{\rm NL}^{\rm loc}}

\newcommand{\Tlm}[1]{T_{\ell m}^{\rm {#1}}}
\newcommand{\hatTlm}[1]{\widehat{T}_{\ell m}^{\rm {#1}}}

\newcommand{\Cl}[1]{C_{\ell}^{\rm {#1}}}

\newcommand{\hatCl}[1]{\widehat{C}_{\ell}^{\rm {#1}}}

\newcommand{\Nl}[1]{N_{\ell}^{\rm {#1}}}

\def\Clpr{\Cl{p}}

\newcommand{\invMpc}{h\, {\rm Mpc}^{-1}\,}
\def\Omm{\Omega_{\rm m}}

\newcommand{\lp}{\left(}
\newcommand{\rp}{\right)}
\newcommand{\lb}{\left[}
\newcommand{\rb}{\right]}

\definecolor{purple}{rgb}{0.78,0.18,0.77}

\newcommand\numberthis{\addtocounter{equation}{1}\tag{\theequation}}

\begin{document}

\title{Cosmic variance mitigation in measurements of the integrated Sachs-Wolfe effect}

\newcommand{\cita}{Canadian Institute for Theoretical Astrophysics, University of Toronto, 60 St.~George Street, Toronto, Canada, M5S 3H8}
\newcommand{\smu}{Department of Physics,
Southern Methodist University, 3215 Daniel Ave, Dallas, TX 75275, U.S.A.}

\newcommand{\kavli}{Kavli Institute for Cosmology, Madingley Road, Cambridge, UK, CB3 0HA}
\newcommand{\damtp}{DAMTP, Centre for Mathematical Sciences, Wilberforce Road, Cambridge, UK, CB3 0WA}
\newcommand{\Kapteyn}{ Kapteyn Astronomical Institute, University of Groningen, P.O. Box 800, 9700 AV Groningen, The Netherlands} 
\newcommand{\VSI}{Van Swinderen Institute for Particle Physics and Gravity,\\ University of Groningen,
Nijenborgh 4, 9747 AG Groningen, The Netherlands}

\author{Simon Foreman}
\affiliation{\cita}

\author{P. Daniel Meerburg}
 \affiliation{\kavli}
  \affiliation{\damtp}
   \affiliation{\Kapteyn}
    \affiliation{\VSI}

\author{Joel Meyers}
\affiliation{\smu}

\author{Alexander van Engelen}
\affiliation{\cita}
\date{\today}

\begin{abstract}
The cosmic microwave background (CMB) is sensitive to the recent phase of accelerated cosmic expansion through the late-time integrated Sachs-Wolfe (ISW) effect, which manifests as secondary temperature fluctuations on large angular scales. However, the large cosmic variance from primary CMB fluctuations limits the usefulness of this effect in constraining dark energy or modified gravity. In this paper, we propose a novel method to separate the ISW signal from the primary signal using gravitational lensing, based on the fact that the ISW signal is, to a good approximation, not gravitationally lensed. We forecast how well we can isolate the ISW signal for different experimental configurations, and discuss various applications, including modified gravity, large-scale CMB anomalies, and measurements of local-type primordial non-Gaussianity. Although not within reach of current experiments, the proposed method is a unique way to remove the cosmic variance of the primary signal, allowing for better CMB-based constraints on late-time phenomena than previously thought possible. 
\end{abstract}

\maketitle

\section{Introduction}

Measurements of cosmic microwave background (CMB) anisotropies have revealed that the universe can be described remarkably effectively by a six-parameter model, with the current energy content dominated by cold dark matter and ``dark energy" consistent with a cosmological constant~\cite{Bennett:2012zja,Akrami:2018vks}. Building on the success of these measurements, future instruments will map CMB temperature anisotropies down to smaller angular scales, while also improving measurements of CMB polarization on all scales~\cite{Ade:2018sbj,Abazajian:2016yjj}. These efforts are motivated in part by the variety of ``secondary effects," sourced during the post-recombination (and in many cases post-reionization) universe, whose impact is most prominent at small angular scales. These include weak gravitational lensing, which shifts power from large to small scales by correlating small-scale fluctuations with background temperature and polarization gradients~\cite{Lewis:2006fu}; the kinetic Sunyaev-Zel'dovich (kSZ) effect, which induces temperature variations when CMB photons scatter off of free electrons with some line of sight velocity~\cite{Sunyaev:1972eq,Sunyaev:1980nv}; and the thermal Sunyaev-Zel'dovich (tSZ) effect, which upscatters CMB photons as they pass through a region of hot gas~\cite{Sunyaev:1972eq}.

A key feature of several of these effects is that they induce couplings between processes acting at different physical scales, such that measurements on small scales carry information about what is happening at larger scales. This feature has been exploited most effectively for gravitational lensing: correlations between small-scale temperature and polarization modes can be optimally combined into an estimator for large-scale modes of the lensing deflection field, enabling reconstruction of projected maps of the massive structures that act as lenses~\cite{Okamoto:2003zw}. Likewise, correlations between small-scale temperature modes and clustered large-scale structure tracers can provide information about the cosmic velocity field on large scales~\cite{Terrana:2016xvc,Deutsch:2017ybc,Cayuso:2018lhv}, and modulation of temperature fluctuations by the velocity field can be used to separate reionization and post-reionization components of the kSZ effect~\cite{Smith:2016lnt,Ferraro:2018izc}. As yet another example, polarization generated by scattering off of free electrons can be used to reconstruct the field of remote CMB quadrupoles~\cite{Deutsch:2017cja}, with possible applications including improved measurements of the mean optical depth to reionization~\cite{Meyers:2017rtf}.
Similar ideas can also be applied to large scale structure surveys: gravitational clustering couples modes of the matter density field in the quasi-linear regime, and these couplings can be used to reconstruct long density modes from shorter-scale measurements, a procedure known as ``tidal reconstruction"~\cite{Pen:2012ft,Zhu:2016esh,Foreman:2018gnv}.

Returning to the case of gravitational lensing, small-scale modes of temperature and polarization will also be correlated with small-scale modes of the projected matter density (which sources lensing), customarily written as a lensing potential field~$\phi$. In Ref.~\cite{Meerburg:2017lfh} (see also~\cite{Cooray:2002ng}), this coupling was exploited to design an estimator for large-scale polarization $E$-modes, which could then be used to precisely measure the mean optical depth to reionization. In Ref.~\cite{Meerburg:2017xga}, it was argued that this same coupling can be used to constrain the amplitude of the intrinsic CMB temperature dipole.

In this paper, we explore a third use of this idea: correlating small-scale temperature and lensing measurements to reconstruct large-scale temperature modes beyond the dipole. This allows a novel way to separate primary temperature fluctuations (sourced around recombination) from fluctuations caused by the integrated Sachs-Wolfe (ISW) effect~\cite{Sachs:1967er,Nishizawa:2014vga}, sourced by time-evolving gravitational potentials during the dark-energy--dominated era at $z\lesssim 1$. ISW fluctuations are not gravitationally lensed in a significant way, and therefore are not correlated with the $\phi$ field in the same way as the primary fluctuations; this implies that, to a good approximation, the reconstructed large-scale modes will only include primary fluctuations, which can then be subtracted from direct observations of large scale temperature fluctuations to isolate the ISW contribution.

Other methods for making ISW maps typically require some assumption about the statistics of the ISW modes, along with external tracers (such as galaxies) that trace the relevant potential wells~\cite{Manzotti:2014kta,Muir:2016veb,Weaverdyck:2017ovf}. Our method requires neither ingredient, and consequently can be used to obtain improved measurements of ISW statistics, that can then be used to constrain various proposals for modified gravity or late-time cosmic growth. Specifically, our method in principle allows more precise measurements of power spectra involving ISW modes than previously thought possible, by effectively eliminating the cosmic variance of the primary CMB on the relevant scales. Reconstructed primary modes can also be used to investigate large-angle ``anomalies," or check that measurements of local-type primordial non-Gaussianity are free of biases from late-time effects.

While intriguing, this reconstruction technique requires unprecedented measurements of the lensing potential down to very small scales, and also requires some level of mitigation of the kSZ component of the temperature spectrum on these scales. Lensing measurements from planned experiments will allow for a noisy reconstruction of the first acoustic peak in the temperature spectrum, while future low-noise, high-resolution experiments~\cite{Nguyen:2017zqu} may enter the regime of useful constraints on ISW or other effects.

This paper is organized as follows. In Sec.~\ref{sec:rec}, we present the estimator for reconstructed modes, state our forecasting assumptions, and quantify the performance of the estimator using a simple parametrization of experimental specifications. In Sec.~\ref{sec:isw}, we explore how well the ISW effect can be measured using reconstructed modes, and discuss possible applications to cosmic variance cancellation and modified gravity. Sec.~\ref{sec:other} comments on further applications to CMB anomalies and measurements of local-type primordial non-Gaussianity. We conclude in Sec.~\ref{sec:conc}.

In this work, we will use the same background cosmology as the Planck 2015 lensing analysis~\cite{Ade:2015zua}: a spatially-flat Lambda cold dark matter model with $\Omega_{\rm b}h^2 = 0.0222$, $\Omm h^2 = 0.1245$, $\Omega_{\nu} h^2 = 0.00064$, $h=0.6712$, $\tau=0.065$, $A_{\rm s}=2.09\times 10^{-9}$, and $n_{\rm s}=0.96$, with the last two parameters fixed at a pivot scale of $k_{\rm pivot}=0.05{\rm Mpc}^{-1}$.  CMB power spectra are computed using CAMB~\cite{Lewis:1999bs}.

\section{Reconstructing large-scale temperature modes}
\label{sec:rec}

\subsection{Estimator}
\label{sec:estimator}

The effect of gravitational lensing on the CMB temperature field $\widetilde{T}(\nhat)$ can be written as a coordinate re-mapping, $\widetilde{T}(\nhat) = T(\nhat + \nabla \phi(\nhat))$, where $T$ is the unlensed temperature field and $\phi$ is the lensing potential, given by a line-of-sight projection of gravitational potentials in direction~$\nhat$. (For now, we will ignore secondary contributions to the temperature field, such that the entire field is sourced at redshift $z_*$ and is lensed by the same~$\phi$.) This induces couplings between different spherical harmonic modes of the temperature field, with the strengths of the couplings related to specific modes of the lensing potential through~\cite{Okamoto:2003zw}
\beq
\widetilde{T}_{\ell_1 m_1}^* = \sum_{\ell_2 m_2 \ell_3 m_3} 
	\Gamma^{\ell_1 \ell_2 \ell_3}_{m_1 m_2 m_3} 
	\phi_{\ell_2 m_2} T_{\ell_3 m_3} \ .
	\label{eq:tonereconstruction}
\eeq
The $\Gamma$ factors above are given by
\beq 
\Gamma^{\ell_1 \ell_2 \ell_3}_{m_1 m_2 m_3} 
	= (-i) e_{\ell_1 \ell_2 \ell_3} I^{ \ell_1 \ell_2 \ell_3}_{m_1 m_2 m_3}\ ,
\eeq
where $e_{\ell_1 \ell_2 \ell_3}$ is equal to unity when the sum $\ell_1 + \ell_2 + \ell_3$ is even and zero when the sum is odd, and 
\begin{align*} 
I^{\ell_1 \ell_2 \ell_3}_{m_1 m_2 m_3} 
	&= \sqrt{\frac{(2\ell_1 + 1)(2\ell_2+1)(2\ell_3+1)}{4\pi}} \\ 
&\quad \times  J_{\ell_1 \ell_2 \ell_3} \lp
	 \begin{matrix} \ell_1 & \ell_2 & \ell_3 \\ m_1 & m_2 & m_3 \end{matrix} 
	 \rp \ , \numberthis
\end{align*}
with
\begin{align*}
J_{\ell_1 \ell_2 \ell_3} &=
	\frac{-\ell_1(\ell_1+1)+\ell_2(\ell_2+1) + \ell_3(\ell_3+1)}{2}  \\ 
&\quad  \times \lp \begin{matrix} \ell_1 & \ell_2 & \ell_3 \\ 0 & 0 & 0 \end{matrix} \rp\ .
	\numberthis
\end{align*}

In standard lensing reconstruction, the couplings in Eq.~\eqref{eq:tonereconstruction} are used to construct an estimator for a given mode of $\phi$ based on a weighted sum of products of observed temperature modes,
\beq
\widehat{\phi}_{LM} = \sum_{\ell_1 m_1 \ell_2 m_2} W^{ \ell_1 \ell_2 L }_{m_1 m_2 M}
	T^{\mathrm{obs}*}_{\ell_1 m_1} T^{\mathrm{obs}*}_{\ell_2 m_2} \ ,
	\label{eq:phiest}
\eeq
with weights $W$ chosen to make the estimator unbiased and of minimum variance under reasonable assumptions about the temperature field. (We use $T^{\mathrm{obs}}$ to denote a field that includes observational noise, while $\widetilde{T}$ denotes a noise-free lensed field.)  The intuition for Eq.~\eqref{eq:phiest} is that a lensing potential mode $\phi_{LM}$ is directly related to the coupling between pairs of temperature modes at different~$\ell$ values, and therefore $\phi_{LM}$ can be reconstructed by correlating many such pairs, typically with $\ell\gg L$. Physically, the coupling is between small-scale temperature anisotropies and larger-scale temperature gradients, and Eq.~\eqref{eq:phiest} can be written to make this manifest.

Eq.~\eqref{eq:tonereconstruction} not only reflects couplings between temperature modes, but it also reflects that temperature modes will be coupled to modes of the lensing potential: $\langle T_{\ell m} \phi_{\ell' m'} \rangle\neq 0$. By analogy with Eq.~\eqref{eq:phiest}, this can be used to construct an estimator of long-wavelength temperature modes:
\beq
\widehat{T}_{LM} = \sum_{\ell_1 m_1 \ell_2 m_2} \bar{W}^{ \ell_1 \ell_2 L }_{m_1 m_2 M}
	T^{\mathrm{obs}*}_{\ell_1 m_1} \phi^{\mathrm{obs}*}_{\ell_2 m_2} \ .
	\label{eq:test}
\eeq
Recall that lensing generates new small-scale fluctuations that depend on the strength of lensing deflections and the magnitude of the background temperature gradient; thus, given a deflection map and a temperature map of small scales, the large-scale gradients can be recovered from this information.

We can fix the weights in Eq.~\eqref{eq:test} by requiring that the estimator be unbiased, $\langle\widehat{T}_{LM} \rangle = T_{LM}$, and minimize its variance, given by (in the Gaussian approximation for $T$ and $\phi$)~\cite{Meerburg:2017lfh} 
\begin{align*}
\mathrm{Var}\left(\widehat{T}_{LM}\right) 
	&= \sum_{\ell_1 m_1 \ell_2 m_2} 
	\bar{W}^{\ell_1 \ell_2 L}_{m_1 m_2 M} \bar{W}^{\ell_1 \ell_2 L *}_{m_1 m_2 M}   \\
&\quad \times  \lp C_{\ell_1}^{TT}+N_{\ell_1}^{TT} \rp 
	\lp C_{\ell_2}^{\phi\phi}+N_{\ell_2}^{\phi\phi}\rp \ ,
	\label{eq:testvar}
	\numberthis
\end{align*}
where $N_\ell^{TT}$ and $N_\ell^{\phi\phi}$ are the noise power spectra on the observed $T$ and $\phi$ fields respectively. These two conditions force
\begin{align*}
\bar{W}^{ \ell_1 \ell_2 L}_{m_1 m_2 M} 
	&= N_L^{\widehat{T} \widehat{T}} 
	\lp \frac{1}{C_{\ell_1}^{TT,\mathrm{res}} + N_{\ell_1}^{TT}} \rp \\ 
&\quad\times \lp \frac{C_{\ell_2}^{\phi\phi} }{C_{\ell_2}^{\phi\phi}+N_{\ell_2}^{\phi\phi}}\rp
	\Gamma^{\ell_1 \ell_2 L *}_{m_1 m_2 M} \ ,
	 \numberthis
\end{align*}
where the reconstruction noise on $T_{LM}$ is given by
\begin{align*}
N_L^{\widehat{T} \widehat{T}} 
	&=  \lb \sum_{\ell_1 \ell_2} e_{\ell_1 \ell_2 L} 
	\frac{(2\ell_1+1)(2\ell_2+1)}{4\pi} 
	\lp J_{\ell_1 \ell_2 L}\rp^2 \right.  \\
&\quad\left. \times   \lp \frac{1}{C_{\ell_1}^{TT,\mathrm{res}} + N_{\ell_1}^{TT}}\rp
	 \lp \frac{(C_{\ell_2}^{\phi\phi})^2 }{C_{\ell_2}^{\phi\phi}+N_{\ell_2}^{\phi\phi}}\rp \rb^{-1}\ .
	 \numberthis
    \label{eq:reconstructionnoise}
\end{align*}
We have replaced the power spectrum of the lensed temperature, $C_{\ell}^{TT}$, by the power spectrum of the temperature field after de-lensing to remove correlations with modes that are not being reconstructed, since this lowers the reconstruction noise (see Ref.~\cite{Meerburg:2017lfh} for further discussion). See App.~\ref{app:flatsky-expressions} for analogous expressions in the flat-sky approximation. 
Similarly to the lensing estimator from Eq.~\eqref{eq:phiest}, the factorizable form of $\bar{W}$ allows this estimator to be rewritten as a product of filtered temperature and $\phi$ maps, which can in practice be efficiently evaluated with fast harmonic transforms~\cite{Okamoto:2003zw}.

So far, we have assumed that the entire temperature field is sourced at the same redshift, and is therefore affected by the same lensing potential (corresponding to a specific redshift weighting of massive structures along the line of sight). However, there are numerous secondary CMB contributions, such as the kinetic and thermal Sunyaev-Zel'dovich effects, inhomogeneous screening due to optical depth fluctuations during reionization, and the integrated Sachs-Wolfe (ISW) effect, that are sourced at lower redshift, and will be subject to different amounts of lensing.

These contributions will only be picked up by the estimator in Eq.~\eqref{eq:test} to the extent that the corresponding lensing potentials correlate with that of the primary CMB; see App.~\ref{app:secondary} for a derivation of the precise impact on~$\widehat{T}_{LM}$. At the large scales we aim to reconstruct, the only relevant effect is ISW, and numerical evaluation of the relevant expression in App.~\ref{app:secondary} reveals that lensing of ISW modes will only bias the resulting $C_L^{\widehat{T}\widehat{T}}$ by less than 0.1\%. Thus, for practical purposes, the $\widehat{T}_{LM}$ allows us to reconstruct modes of the {\em primary} CMB on their own, and late-time or systematic effects can be isolated by comparing reconstructed and directly-measured large-scale modes.

Note that this procedure is distinct from delensing, which uses  a $\phi$ map to recover the unlensed temperature by inverting the pixel re-mapping induced by lensing. Rather, our procedure makes use of the lensing of an observed map on small scales to recover the large-scale fluctuations whose deflection generated the new small-scale fluctuations.

\subsection{Forecasting assumptions}
\label{sec:forecasts}

In our forecasts, we will consider idealized experiments that can accomplish cosmic-variance-limited measurements of the small-scale temperature and lensing potential fields up to some $\lmax$ (i.e.\ $N_\ell^{TT}=N_\ell^{\phi\phi}=0$ for $\ell\leq\lmax$ and infinity for $\ell>\lmax$), and assess the performance of the reconstruction procedure as $\lmax$ is varied. 
(This is a reasonable approximation for the temperature measurements, because the signal to noise typically undergoes a sharp transition over a narrow range of scales.) If lensing is reconstructed internally from quadratic estimators applied to CMB temperature and polarization maps, the lensing noise will depend on the signal to noise in those maps; the lensing map is not expected to be signal-dominated on the same scales as temperature and polarization.
One could imagine specifying separate $\lmax$ values for $T$ and $\phi$; if both are obtained from the same experiment, then we will have $\lmax^\phi<\lmax^T$, and temperature modes with $\ell>\lmax^\phi+L_{\rm max}$ will not enter the reconstruction. Thus, for low $L_{\rm max}$ (no more than a few hundred), having~$\lmax^T$ much greater than $\lmax^\phi$ will make little difference, so we use the same~$\lmax$ for both. 

Maps of the lensing potential can also be estimated from galaxy surveys, observations of the cosmic infrared background, or other external tracers; to date, these have mainly been discussed in the context of delensing the CMB~\cite{Smith:2010gu,Sherwin:2015baa,Larsen:2016wpa,Yu:2017djs}.  These maps are unlikely to capture the contributions to the lensing potential at very high redshift, but for sufficiently low shot noise in the tracer sample, they can in principle carry information about lenses on smaller scales ($\ell\gtrsim 1000$) than can be obtained from internal lensing reconstruction for currently funded experiments.

The kSZ effect has a significant contribution to the temperature power spectrum at small scales, exceeding the primary CMB power for $\ell\gtrsim 3000$~\cite{Shaw:2011sy}, and, unlike other foregrounds, cannot be cleaned from CMB maps using multifrequency information. The associated variance in observed temperature maps is a limiting factor on the precision of long mode reconstruction when $\lmax\gtrsim 4000$, so we will consider the impact of including this contribution or not. The latter is motivated by the idea that some degree of ``kSZ cleaning" may be possible in the future (see e.g.~\cite{Smith:2016lnt,Ferraro:2018izc} for ideas along these lines), in which case the reconstruction's precision would fall somewhere in between the two cases we consider. When including kSZ power, we use the model from Ref.~\cite{Shaw:2011sy}. The numerical results from this model are consistent with current upper limits on the kSZ power spectrum~\cite{George:2014oba}, and also with recent indications of a short, late period of reionization~\cite{Kulkarni:2018erh}, which will drive the corresponding kSZ contribution to be smaller than the post-reionization contribution.

We will also consider CMB-S4~\cite{Abazajian:2016yjj},  assuming a $1'.4$ beam and a white noise level of $1\,\mu$K-arcmin in temperature, and a minimum-variance combination of quadratic~$\phi$ estimators involving $T$, $E$, and $B$ modes, including the improvement that comes from iterative $EB$ reconstruction~\cite{Smith:2010gu}. 
This setup can accomplish signal-dominated measurements of $T$ for $\ell\lesssim 3500$ and $\phi$ for $\ell\lesssim 1000$, but due to the mild scaling of the~$\phi$ noise with $\ell$, the resulting reconstruction noise on long $T$ modes roughly matches that of a cosmic-variance--limited experiment with $\lmax\approx 2000$ (see App.~\ref{app:flatsky-noise} for further discussion).

The variance of the long-mode estimator will also receive contributions beyond Eq.~\eqref{eq:testvar}, analogous to what is often called ``$N^1$ bias" in CMB lensing~\cite{Kesden:2003cc}, but these are far subleading at the  low $L$ values we work at, and can safely be neglected.

\subsection{Results for primary modes}
\label{sec:primary}

\begin{figure}[t]
\includegraphics[width=\columnwidth]{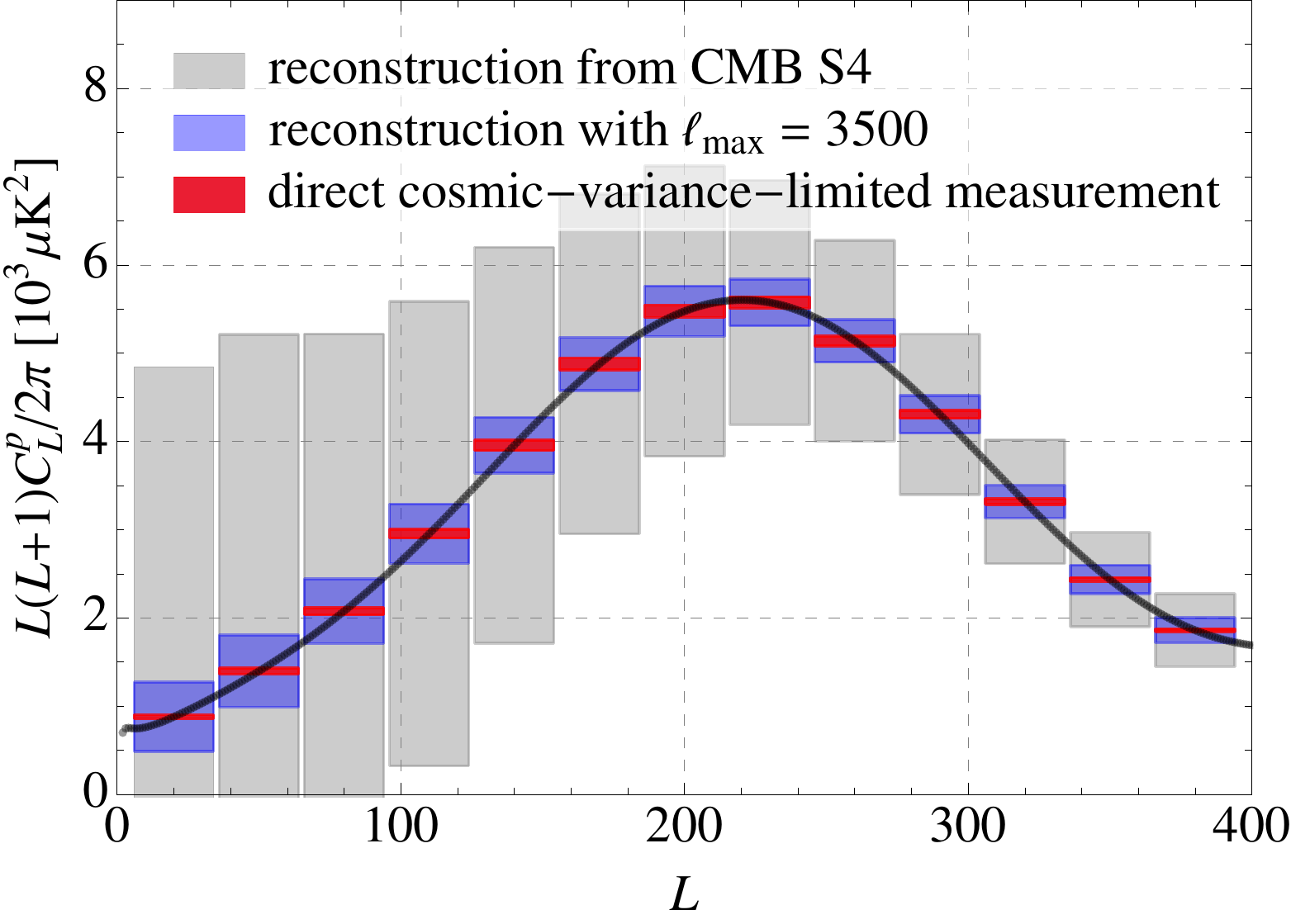}
\caption{Expected errorbars on the primary CMB temperature power spectrum, as bandpowers with $\Delta L=30$, either reconstructed with the estimator in Eq.~\eqref{eq:test} using small-scale measurements from CMB-S4 ({\em grey}), reconstructed using cosmic-variance--limited $T$ and $\phi$ modes up to $\lmax=3500$ ({\em blue}), or directly measured by e.g.~Planck ({\em red}). CMB-S4 can obtain a high-significance reconstruction of the entire first peak, but more ambitious measurements at small scales will be required to approach the precision of direct measurements at large scales.
}
\label{fig:clttp-binned-errorbars}
\end{figure}

In Fig.~\ref{fig:clttp-binned-errorbars}, we show the expected errorbars on a reconstruction of the primary temperature power spectrum, plotted as bandpowers with $\Delta L=30$, using the estimator in Eq.~\eqref{eq:test}, for two representative cases: reconstruction with CMB-S4, and reconstruction using noise-free~$T$ and~$\phi$ modes up to $\lmax=3500$. 
We also show errorbars for a cosmic-variance--limited direct measurement of the power spectrum, representative of Planck's existing measurement. We see that a high-significance reconstruction of the entire first acoustic peak will be possible with CMB-S4, while the availability of smaller-scale measurements of temperature and lensing (or high-fidelity proxies of the lensing potential) would allow for the reconstruction to approach the cosmic variance limit of a direct measurement.

\begin{figure}[t]
\includegraphics[width=\columnwidth, trim = 0 0 25 25]{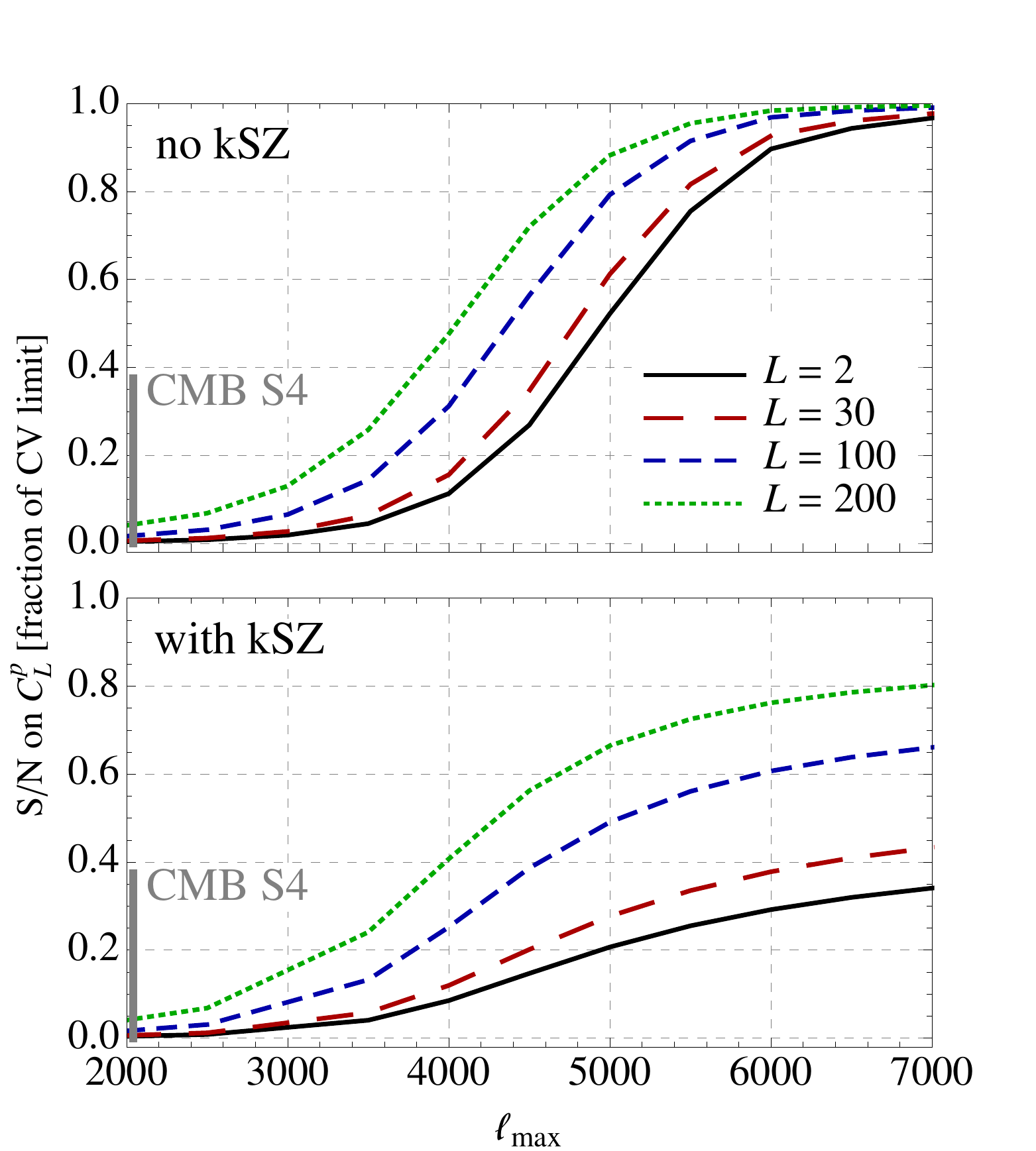}
\caption{
Expected signal to noise on the reconstructed primary power spectrum $C_L^{\rm p}$ at several $L$ values, normalized to the cosmic variance limit at the same $L$, if noise-free small-scale $T$ and $\phi$ modes up to $\lmax$ are used in the reconstruction. If the kSZ contribution can be perfectly cleaned from the small-scale $T$ modes, $C_L^{\rm p}$ can be reconstructed at near--cosmic-variance precision for $\lmax\gtrsim 5000$ ({\em upper panel}), while if kSZ cannot be cleaned at all, the reconstructed modes will have more limited precision ({\em lower panel}). CMB-S4 is roughly equivalent to the $\lmax\approx 2000$ case (see main text for discussion).
}
\label{fig:clttp-sn-ratio}
\end{figure}

In Fig.~\ref{fig:clttp-sn-ratio}, we plot the expected signal to noise on the reconstructed primary power spectrum at several~$L$ values, normalized to the cosmic variance limit at the same~$L$,
\beq
\frac{ \lp \text{S/N} \rp_{\rm rec.} }{ \lp \text{S/N} \rp_{\rm CV} }
	= \frac{1}{ 1+N_L^{\widehat{T}\widehat{T}}/C_L^{\rm p} }\ ,
\eeq
both with and without including kSZ power in temperature. Since the reconstruction noise $N_L^{\widehat{T}\widehat{T}}$ is roughly white when multiplied by $L^2$, the differences between the plotted curves reflect the shape of $L^2 C_L^{\rm p}$; we have chosen $L$ values that span the range between the minimum ($L\approx 10$) and maximum ($L\approx 200$) values of $L^2 C_L^{\rm p}$. For perfect kSZ cleaning and $\lmax\gtrsim 6000$, the noise on the reconstructed spectrum is within 10\% of the cosmic variance limit, while without any cleaning, the improvement with $\lmax$ is much slower.

\begin{figure}[t]
\includegraphics[width=\columnwidth]{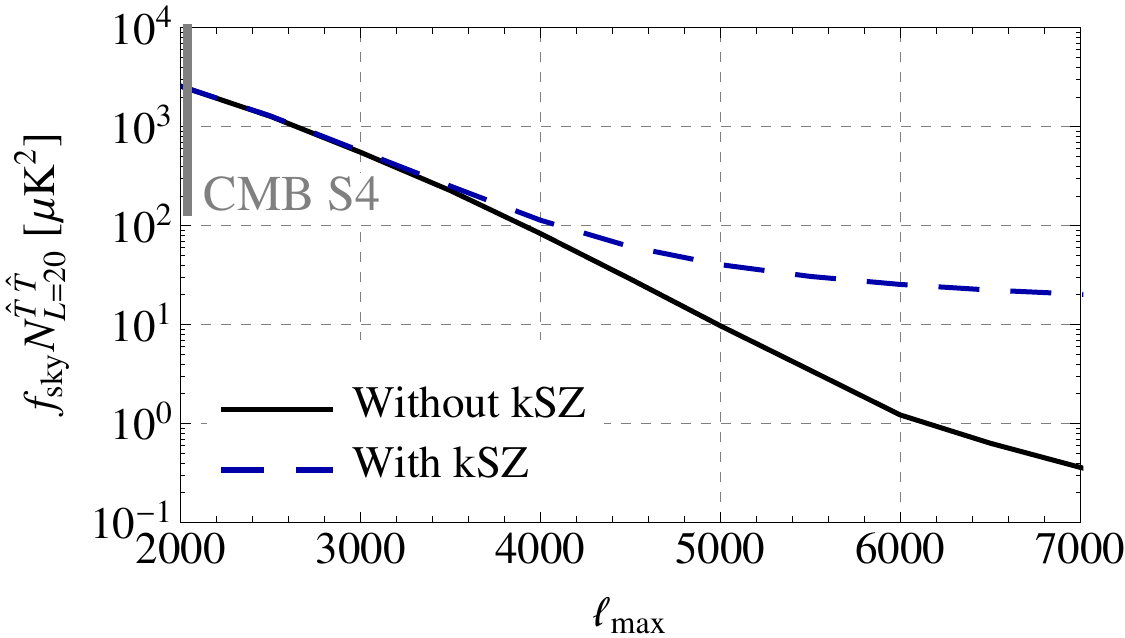}
\caption{
The noise per reconstructed mode, in absolute temperature units and evaluated at $L=20$, for reconstruction using full-sky measurements with the specified $\lmax$. These curves can be rescaled to lower sky fractions by dividing by~$f_{\rm sky}$, or to other $L$ values using the fact that $L^2 N_L^{\widehat{T}\widehat{T}}$ is roughly constant with $L$.
}
\label{fig:nltt}
\end{figure}

Fig.~\ref{fig:clttp-sn-ratio} also shows that, for perfect kSZ cleaning, individual reconstructed temperature modes become signal-dominated for $\lmax\gtrsim 4000$ to $5000$ depending on $L$. If kSZ cannot be cleaned, the reconstruction of modes with $L\lesssim 50$ will always be noise-dominated. Fig.~\ref{fig:nltt} shows a complementary view of the noise per mode; specifically, we plot $f_{\rm sky} N_L^{\widehat{T}\widehat{T}}$ at $L=20$ in absolute temperature units. This can be related to other $L$ values using the fact that $L^2 N_L^{\widehat{T}\widehat{T}}$ is roughly constant with $L$.

\section{Measuring the Integrated Sachs-Wolfe Effect}
\label{sec:isw}

\subsection{Auto and cross spectra}
\label{sec:iswspectra}

The ISW effect\footnote{The ISW effect occurs both at ``early" times just after recombination, due to radiation's non-negligible contribution to the cosmic energy budget at those times, and at ``late" times, due to the onset of dark energy domination around $z\sim 1$. In this paper, we use ``ISW" to refer exclusively to the late-time effect, since the early-time effect will be reconstructed by the estimator in Eq.~\eqref{eq:test} just like the other primary contributions.}~\cite{Sachs:1967er,Nishizawa:2014vga} is produced when gravitational potentials evolve in time, causing a net blueshift or redshift of photons that pass through potentials that are large enough. All such blueshifts or redshifts along each line of sight affect the photon temperature observed in that direction (e.g.~\cite{Kim:2013nea}):
\beq
\left. \frac{\Delta T}{T_0} \right|_{\rm ISW}(\nhat) 
	= \int_0^{\chi_*} d\chi \lb \dot{\Psi}-\dot{\Phi} \rb (\chi\nhat,\eta_0-\chi)\ ,
	\label{eq:iswfull}
\eeq
where $\Psi$ and $\Phi$ are the Newtonian potential and curvature perturbation, respectively, in conventions where the Newtonian-gauge metric components are $g_{00}=-a^2(1+2\Psi)$ and $g_{ii}=a^2(1+2\Phi)$; $\chi$ is the comoving line-of-sight distance; $\chi_*$ is the distance to the last-scattering surface; dots denote derivatives with respect to conformal time $\eta$; and $\eta_0$ is the conformal time at $z=0$. If $\Phi=-\Psi$ (which is true in general relativity in the absence of anisotropic stress), we can use the Poisson equation to rewrite this as
\begin{align*}
\left. \frac{\Delta T}{T_0} \right|_{\rm ISW}(\nhat) 
	&=  \frac{3\Omm H_0^2}{c^2} 
	\int_0^{\infty}  dz  \lb f(z)-1 \rb \\
&\qquad\qquad\qquad\quad \times \d^{-2} \delta(\chi\hvn;z[\chi])\ ,
	\numberthis
	\label{eq:iswlinear}
\end{align*}
where $f(z) \equiv \d\log D/\d\log a$ is the logarithmic growth rate and $\d^{-2}$ is understood to act as $-k^{-2}$ in Fourier space. This can then be used to derive expressions for various angular auto- and cross-spectra (e.g.~\cite{Muir:2016veb}). 

Note that Eq.~\eqref{eq:iswfull} includes the fully nonlinear $\Phi$ and $\Psi$, and therefore incorporates the nonlinear (Rees-Sciama) ISW effect, while Eq.~\eqref{eq:iswlinear} is only valid in the linear regime. For our forecasting purposes, Eq.~\eqref{eq:iswlinear} is sufficient, but the ISW map $\hatTlm{ISW|WF}$ that we will define later will also include nonlinear effects, to the extent that they matter on the relevant scales.

The power spectrum of the ISW effect, $\Cl{ISW}$, is notoriously difficult to isolate in CMB measurements. One could imagine doing so by subtracting the best-fit theoretical primary spectrum from the measured spectrum,
\beq
\hatCl{ISW|sub} = \Cl{obs}-\Clpr\ ,
\label{eq:iswautosub}
\eeq
but the uncertainty of this estimate will be  dominated by the primary contribution, 
\begin{align}
\label{eq:sigiswautosub}
\sigma\!\lp \hatCl{ISW|sub} \rp &= \sqrt{\frac{2}{(2\ell+1)f_{\rm sky}}} \Cl{obs} \\
&= \sqrt{\frac{2}{(2\ell+1)f_{\rm sky}}} \lp \Cl{ISW} + \Clpr + \Nl{obs} \rp\ .
\end{align}
Even for noise-free temperature measurements, the maximum cumulative signal to noise on the entire $\Cl{ISW}$ spectrum is about $1\sigma$ using this method. One can do better on the cross spectrum between ISW and a tracer $X$ of low-redshift gravitational potentials, $C_\ell^{X\times{\rm ISW}}$. The direct cross-correlation will have uncertainty
\begin{align*}
\label{eq:sigxisw}
\numberthis
\sigma\!\lp \widehat{C}_\ell^{X\times{\rm ISW|dir}} \rp  
	& = \frac{1}{\sqrt{(2\ell+1)f_{\rm sky}}} \\
& \times \lb \lp C_\ell^{X\times{\rm ISW}} \rp^2 
 + \lp C_\ell^X+N_\ell^X \rp \Cl{obs} \rb^{1/2} ,
\end{align*}
with the primary contribution to $\Cl{obs}$ limiting the cumulative signal to noise to no more than $\sim$$7\sigma$ (if~$X$ is a perfect tracer)~\cite{Crittenden:1995ak,Afshordi:2004kz,Ferraro:2014msa}. The latest analyses have reached $\sim$$5\sigma$ using galaxies or quasars as tracers~\cite{Stolzner:2017ged} and $\sim$$3\sigma$ using CMB lensing~\cite{Ade:2015dva}.

The estimator from Sec.~\ref{sec:rec} allows us to improve upon these measurements by implementing the subtraction in Eq.~\eqref{eq:iswautosub} {\em mode by mode} rather than in the power spectrum. We can account for noise in the reconstructed primary modes by Wiener-filtering them before subtraction,
\beq
\label{eq:tiswwf}
\hatTlm{ISW|WF} = \Tlm{obs} - \frac{\Cl{p}}{C_\ell^{\widehat{T}\widehat{T}}} \hatTlm \ ,
\eeq
where $C_\ell^{\widehat{T}\widehat{T}} = \Cl{p}+ N_\ell^{\widehat{T}\widehat{T}}$.
The power spectrum of these modes is then
\beq
\hatCl{ISW|WF} = \Cl{ISW} + \Nl{obs} 
	+  \frac{\Cl{p} N_\ell^{\widehat{T}\widehat{T}}}{\Cl{p}+ N_\ell^{\widehat{T}\widehat{T}}} \ ,
\eeq
with uncertainty given by Eq.~\eqref{eq:sigiswautosub} with the substitution $\Cl{obs}\to \hatCl{ISW|WF}$. The uncertainty on the cross spectrum between these modes and a tracer $X$ is given by Eq.~\eqref{eq:sigxisw} with the same substitution. It is clear from these expressions that for low reconstruction noise, the primary contribution to the uncertainty disappears (effectively removing the cosmic variance from the primary modes), while for high reconstruction noise, this procedure is equivalent to the case with no reconstruction.

\begin{figure}
\includegraphics[width=\columnwidth, trim = 0 0 25 25]{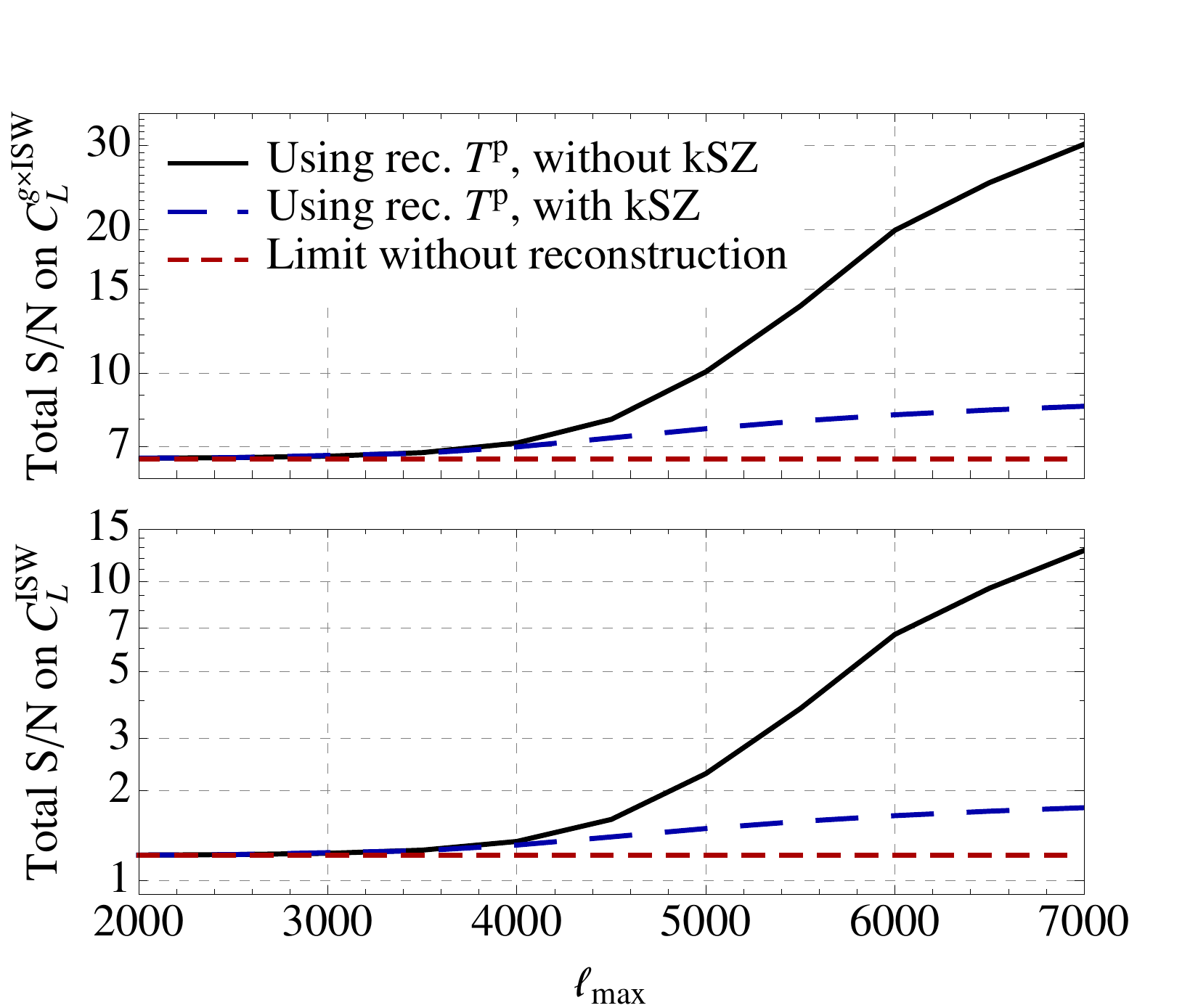}
\caption{
Total signal to noise on ISW cross (with a perfect galaxy tracer; {\em upper panel}) or auto spectra ({\em lower panel}), using an ISW map constructed from the difference between directly-measured large-scale temperature modes and Wiener-filtered reconstructed primary modes [see Eq.~\eqref{eq:tiswwf}]. As in the previous figures, $\lmax$ denotes the maximum multipole of the~$T$ and~$\phi$ modes used for reconstruction. We also show the S/N in the absence of reconstruction. If kSZ can be entirely cleaned at small scales, reconstruction can significantly improve the S/N on both the cross-  and auto-spectra if $\lmax$ is sufficiently high. Note that all curves assume complete sky coverage, and should be scaled by $f_{\rm sky}^{1/2}$ for a smaller sky area.
}
\label{fig:iswspectra}
\end{figure}

In Fig.~\ref{fig:iswspectra}, we show the cumulative signal to noise%
\footnote{
We compute the signal to noise from $L_{\rm min}=2$ to $L_{\rm max}=100$, ignoring off-diagonal covariance between different power spectrum multipoles. At $L=100$, the relevant auto and cross spectra receive no more than 20\% of their value from $z\lesssim 0.7$, which at $L=100$ translates to $k\approx 100/\chi(z=0.7)\approx 0.06\invMpc$. At this scale and redshift, off-diagonal covariance has negligible impact on the information content of the matter power spectrum~\cite{Repp:2015jja,Pan:2016dtf}, so we are justified in using a diagonal covariance in our forecasts.
} 
on a measurement of ISW cross- (with galaxies that form a perfect ISW tracer) or auto-spectra with reconstruction, as a function of the $\lmax$ used for the reconstruction. With perfect kSZ cleaning, substantial improvements over the standard cross-correlation analysis are possible, with signal to noise reaching $\sim$30 for reconstruction with $\lmax=7000$, corresponding to a $30\sigma$ detection of dark energy via its ISW signature. Without any kSZ cleaning, the improvements on the cross spectrum are more modest. For the auto spectrum, some degree of kSZ cleaning will be required to achieve a significant detection, but in optimistic cases, a measurement at greater than $5\sigma$ can be attained. To the best of our knowledge, this is the only method that can enable a confident detection of the ISW auto spectrum.

One could also consider using measurements of large-scale $E$-mode polarization as a proxy for primary temperature modes, exploiting the nonzero $TE$ correlation at large scales to produce a Wiener-filtered primary $T$ map that can then be subtracted from direct measurements as in Eq.~\eqref{eq:tiswwf}. This method was discussed in Ref.~\cite{2009MNRAS.395.1837F} and applied to data in Refs.~\cite{Liu:2010re,Giannantonio:2012aa}, where it yielded roughly 15\% improvements on signal to noise in the ISW cross spectrum. The maximum possible gain from this method is roughly 20\%, due to imperfect $TE$ correlation at the relevant scales~\cite{2009MNRAS.395.1837F}, and therefore our reconstruction technique will perform better for a sufficiently high~$\lmax$.

\subsection{Cosmic variance cancellation}

One can ask whether a combined analysis involving galaxy-ISW cross spectra and galaxy clustering auto spectra could be used to obtain cosmic-variance--free tomographic measurements of the growth function $f(z)$, similar to what has been proposed for constraining local-type non-Gaussianity and the amplitude of matter clustering with galaxy clustering and CMB lensing~\cite{Seljak:2008xr,Schmittfull:2017ffw,Yu:2018tem}. We investigate this using a simple Fisher forecast, considering an ideal case in which all other relevant parameters (linear bias and cosmology) are perfectly known. We scale $1-f(z)$ by a separate constant factor $\alpha$ within each of a series of redshift bins, and compute the relative precision that can be obtained on each factor. For a bin bounded by $z_{\rm min}$ and $z_{\rm max}$, the Fisher matrix element for the corresponding $\alpha$ is given by~\cite{Schmittfull:2017ffw}
\begin{align*}
F_{\alpha\alpha} &= \sum_{L_{\rm min}}^{L_{\rm max}}
	\frac{(2L+1)f_{\rm sky}}{\lp 1-r_L^2 \rp^2}
	\lb \lp \frac{C_L^{\rm ISW,bin}}{\widehat{C}_L^{\rm ISW}} - r_L^2 \rp^2 \right. \\
&\qquad\qquad\qquad\qquad\qquad\qquad \left.+\, 2r_L^2\lp 1-r_L^2 \rp \rb\ ,
	\numberthis
\end{align*}
where $r_L \equiv C_L^{{\rm g}\times{\rm ISW,bin}}(\widehat{C}_L^{\rm gg,bin} \widehat{C}_L^{\rm ISW})^{-1/2}$ and ``bin" superscripts refer to the contribution to the relevant spectrum from the specified bin. The relative precision obtainable on $\alpha$ is then given by $(F_{\alpha\alpha})^{-1/2}$.

In the limit $r_L\to 1$, the resulting precision on $\alpha$ becomes arbitrarily high, but in practice, the precision is limited by several factors, including shot noise in the galaxy sample, imperfect overlap between the galaxy clustering and ISW redshift kernels, and reconstruction noise on the ISW modes used in the analysis. (That is not to mention imperfect knowledge of other parameters, in which case we would need to estimate the marginalized errorbar using $([F^{-1}]_{\alpha\alpha})^{1/2}$.) Even if these obstacles were not present---for a perfect, shot-noise-free ISW tracer, and with zero ISW reconstruction noise---we obtain $r_L=(C_L^{\rm ISW,bin}/C_L^{\rm ISW})^{1/2}$, and therefore the precision obtainable on $\alpha$ will be limited by the width of the chosen redshift bins.

As a concrete example, we have considered how well $\alpha$ could be constrained in redshift bins with $\Delta z = 0.1$, by using cross- and auto-correlations between reconstructed ISW modes and a perfect galaxy tracer. We find that even for $\lmax=7000$ and perfect kSZ cleaning, the expected constraints on $\alpha$ will be no better than a factor of a few higher than those expected on $f(z)\sigma_8(z)$ from redshift space distortions in DESI~\cite{Aghamousa:2016zmz}. Thus, direct constraints on $f(z)$ using this method are unlikely to be competitive for the foreseeable future.

\subsection{Modified gravity}

The ISW effect can provide interesting cosmological constraints on its own. In particular, while redshift space distortions probe structure growth through the relationship between velocities and the gravitational potential~$\Psi$, the ISW effect probes the evolution of the full Weyl potential $\Psi-\Phi$. This fact makes the ISW effect an especially strong discriminator between modified gravity theories that change the behavior of the Weyl potential or its relationship to the matter density (e.g.~\cite{Lue:2003ky,Song:2006jk,DiValentino:2012yg,Renk:2016olm}). Measurements of the ISW effect have been used to place constraints on $f(R)$ gravity~\cite{Giannantonio:2009gi}, DGP gravity~\cite{Fang:2008kc,Lombriser:2009xg}, and Horndeski models~\cite{Kreisch:2017uet}, as well as completely rule out cubic Galileon models as an explanation for dark energy~\cite{Renk:2017rzu}.

More precise ISW measurements, enabled by the method in this paper, will be able to continue that trend. To cite a specific example, Ref.~\cite{Enander:2015vja} has computed the ISW effect within a cosmologically-viable branch of ghost-free massive bigravity. They find the ISW auto spectrum to be roughly a factor of 4 higher than in $\Lambda$CDM, implying that a $\sim$$4\sigma$ measurement of this spectrum could rule out this theory at $\sim$$3\sigma$. Fig.~\ref{fig:iswspectra} shows that this could be achieved with $\lmax\sim 5700$ with perfect kSZ cleaning; while this is an ambitious goal to realize experimentally,
use of the auto spectrum for this purpose would be impossible without any long-mode reconstruction. Ref.~\cite{Enander:2015vja} also computes the cross spectrum between ISW and galaxies with redshift distribution similar to the WISE survey (mimicking the measurements from Ref.~\cite{Ferraro:2014msa}), finding an amplitude roughly 1.5 times higher than in $\Lambda$CDM, and reconstruction would also help to obtain more precise measurements of this cross-correlation.

Furthermore, since the ISW effect depends on the time evolution of gravitational potentials, correlations of ISW modes with other tracer fields effectively probe unequal-time correlations of the cosmic density field, which are otherwise difficult to access in observations. In general relativity and for Gaussian initial conditions, these correlations must obey consistency relations~\cite{Rizzo:2016akm,Rizzo:2017zow} in the squeezed limit, similar to those that apply for equal-time density correlations~\cite{Peloso:2013zw,Creminelli:2013mca,Kehagias:2013yd}. Deviations from these relations would signal a violation of the equivalence principle or the presence of primordial non-Gaussianity~\cite{Kehagias:2013rpa,Creminelli:2013nua,Rizzo:2017zow}, and could potentially be checked by correlating reconstructed ISW modes with two or more modes of another tracer.

\section{Other applications}
\label{sec:other}

\subsection{Testing CMB anomalies}

A number of somewhat surprising statistical features have been identified in CMB temperature maps, ranging from low values of the pixel-space correlation function at large angular separations, to a hemispherical power asymmetry, to a large cold spot in the Southern hemisphere (e.g.~\cite{Ade:2015hxq,Aghanim:2016sns}; see Ref.~\cite{Schwarz:2015cma} for a review). Many of these features reside at large scales that are accessible to our reconstruction procedure. Therefore, comparing reconstructed primary CMB modes with direct measurements at these scales would allow us to investigate whether these features have a primordial origin, or might be due to late-time effects or systematics (see Ref.~\cite{Ballardini:2018noo} for a related approach that makes use of an externally-estimated ISW map rather than reconstructed primary modes).

However, the low statistical significance of these features implies that the reconstruction procedure will have to reach high precision to verify them with any confidence. For example, the lack of large-angle correlations and low map variance can be traced to a deficit of power for $\ell<30$, which affects cosmological parameter values by 1-2$\sigma$~\cite{Aghanim:2016sns,Addison:2015wyg}, and also affects constraints on primordial scalar and tensor power spectra (e.g.~\cite{Akrami:2018odb}). The measured multipoles at $\ell<30$ have an overall amplitude that is roughly 2$\sigma$ lower than that derived from the best fit over the entire measured spectrum~\cite{Aghanim:2016sns}. This requires us to reconstruct modes at very close to cosmic variance precision to detect this power deficit; Fig.~\ref{fig:clttp-sn-ratio} shows that this will in turn require both some degree of kSZ cleaning and a very high effective $\lmax$ (at least 5500 with perfect kSZ removal, or higher with a lower cleaning efficiency).

One can also consider the Cold Spot, and ask whether reconstructed modes could be used to test its presence in the primary CMB (as opposed to generation by late-time effects\footnote{Note that if the Cold Spot results from the presence of some evolving structure in the low redshift Universe, it is also possible to detect the gravitational lensing due to that structure \cite{Das:2008es}.} \cite{Cruz:2008sb}). 
An efficient way to do this is to cross-correlate the reconstructed modes with the directly observed modes within the relevant patch of sky: the corresponding cross power spectrum will vanish if the Cold Spot has a non-primordial origin. Taking this as our null hypothesis, the total significance with which we can distinguish this case and a completely primordial origin (in which case the cross power spectrum will equal the primary power spectrum for a perfect reconstruction) is given by
\beq
\frac{\rm S}{\rm N}
	= \lb \sum_{L=L_{\rm min}}^{L_{\rm max}} (2L+1)f_{\rm sky} 
	\frac{C_L^p}{N_\ell^{\widehat{T}\widehat{T}}} \rb^{1/2}\ .
	\numberthis
\eeq
We take $L_{\rm min}=10$ and $L_{\rm max}=50$, corresponding roughly to the scales covered by the wavelet function used to detect the Cold Spot in Ref.~\cite{Ade:2015hxq}. We find that the two origins for the Cold Spot can be distinguished at $\gtrsim 3$$\sigma$ for reconstruction with $\lmax\gtrsim 4500$ and perfect kSZ removal, or $\lmax\gtrsim 5300$ with no kSZ removal.

These examples clearly set very ambitious targets for small-scale measurements, particularly of lensing, that can then be used for reconstruction. However, it is worth noting that reconstruction could in principle be accomplished {\em using only ground-based CMB measurements}. Low-$\ell$ modes are typically inaccessible from the ground due to atmospheric noise or other systematics, while high-$\ell$ modes are more naturally measured from the ground; using the latter to reconstruct the former would act as an interesting complement to direct (space-based) measurements of low-$\ell$ information.

\subsection{Eliminating bias on primordial non-Gaussianity measurements}

A key science target for future cosmological measurements is to detect or constrain non-Gaussianity of the primordial perturbations that acted as seeds for all large-scale structure we observe at recent times (e.g.~\cite{Alvarez:2014vva}). Of particular interest is so-called local-type non-Gaussianity, typically quantified using the parameter $\fnllocal$. A measurement of $\fnllocal\gtrsim 1$ would provide strong evidence against single-field models of cosmic inflation; the current best limit has $\sigma(\fnllocal) \approx 5$~\cite{Ade:2015ava}, and using the primary CMB from CMB-S4 is expected to reduce this to~$\sim$$2$, largely thanks to improved polarization measurements~\cite{Abazajian:2016yjj}, or $\sim$$4$ if only temperature is used~\cite{Hill:2018ypf}.

CMB-based constraints on $\fnllocal$ are driven by the amplitude of temperature and polarization bispectra in the squeezed limit ($\ell_1\ll \ell_2,\ell_3$ or permutations). The cross-correlation between gravitational lensing and the ISW effect also produces a bispectrum in this configuration, since the lensing potential $\phi$ is estimated using two short modes of temperature or polarization, while ISW contributes to long modes of temperature. This results in a bias on an estimate of $\fnllocal$ (equal to $7.6$ for Planck~\cite{Ade:2015ava}) that must be subtracted in order to access the primordial value.

It has recently been pointed out~\cite{Hill:2018ypf} that other similar biases exist, arising from bispectra between contributions to the CMB temperature from the cosmic infrared background (CIB) or the tSZ and kSZ effects. The CIB and tSZ contributions can in principle be cleaned from temperature maps due to their non-blackbody spectral shapes, but this cleaning must be very efficient to avoid sizable residual biases on $\fnllocal$.

On the other hand, if the squeezed temperature mode is taken from the reconstruction method in this paper, its bispectrum with two short modes will be much less contaminated by these late-time correlations. Contaminations arising from the ISW effect (for example, the ISW-tSZ-tSZ bispectrum) will essentially be absent, since the residual ISW contribution to the reconstructed mode will only be at the percent level. Other terms discussed in Ref.~\cite{Hill:2018ypf}  will likewise be strongly reduced.

Of course, the reconstructed modes will come with reduced precision compared to direct measurements of those modes, and this will degrade the statistical uncertainty on $\fnllocal$. For CMB-S4, the noise on the reconstructed modes will be too high to make them useful for this purpose. For future surveys with lower reconstruction noise, however, these modes will be useful as a check of the multi-frequency cleaning and de-biasing procedures that must be used in the main $\fnllocal$ analysis. Consistency between the $\fnllocal$ constraints obtained with and without reconstruction would provide further confidence that the constraints are robust to late-time biases.

One possible caveat is the presence of other biases even when a reconstructed long mode is used: schematically, since $T_{\rm long}^{\rm rec.} \sim \la \phi_{\rm short} T_{\rm short} \ra$, the bispectrum $\la T_{\rm long}^{\rm rec.} T_{\rm short}T_{\rm short}  \ra$ will contain four-point correlations between one short mode of $\phi$ and three short modes of $T$, along with five-point correlations between five short modes of~$T$ if~$\phi$ is itself obtained from a quadratic estimator in~$T$. Evaluating the possible terms involving CIB, tSZ, and kSZ is beyond the scope of this work, but could be accomplished using either the analytical methods of Ref.~\cite{Hill:2018ypf} or appropriately correlated simulations~\cite{vanEngelen:2013rla,stein-peak-patch}.

\section{Conclusions}
\label{sec:conc}

In this paper, we have presented an estimator that uses small-scale measurements of CMB temperature and lensing to reconstruct large-scale information in the primary CMB, similarly to how large-scale lensing information can be reconstructed from correlations of small-scale temperature or polarization modes. We have discussed a number of applications, focusing in particular on how this estimator can allow for improved measurements of two-point statistics involving the integrated Sachs-Wolfe effect, circumventing the standard lore that such measurements are fundamentally limited by the cosmic variance of the primary CMB. These improvements would be particularly helpful for constraining different modified gravity scenarios, since the ISW effect has provided a multitude of constraints on such theories~\cite{Giannantonio:2009gi,Fang:2008kc,Lombriser:2009xg,Kreisch:2017uet,Renk:2017rzu} even with the current, somewhat limited, precision.

Ambitious criteria must be met for these improvements to be realized in practice: high-resolution lensing maps (resolving modes up to $\ell\sim 4000$ at least) and some method to clean the small-scale kSZ signal from temperature maps used in the estimator. We leave the second item to future work, but note that the first item is a primary goal of a recent proposal for low-noise, high-resolution CMB observations, motivated by different dark matter scenarios that could be tested using such observations. The fiducial case considered by Ref.~\cite{Nguyen:2017zqu}, of a 4000$\,$deg$^2$ survey with $T$ and $\phi$ measured up to $\lmax\sim 35000$, would likely have the statistical power to incur substantial improvements on ISW measurements through our reconstruction procedure, but would again be limited by kSZ at small scales, and it is an open question to what extent this can be mitigated.
We note that external tracers, such as the cosmic infrared background, could also be used for the lensing field, but these would also need to be mapped to sufficiently small scales in order to reduce the shot noise contribution to their power spectra.

With additional applications to large-scale anomalies and primordial non-Gaussianity, the reconstruction procedure in this paper would be a novel use of small-scale CMB temperature and lensing measurements, and provides additional motivation for future experiments to push further into the small-scale, low-noise regime.

\section*{Acknowledgments}
\noindent
We would like to thank  Anthony Challinor, Harry Desmond, Eiichiro Komatsu, Antony Lewis, Mathew Madhavacheril, Neelima Sehgal, and David Spergel for useful discussions.  P.~D.~M.\ thanks CITA for hospitality while this work was being completed. P.~D.~M.\ acknowledges support from Senior Kavli
Institute Fellowships at the University of Cambridge and the Netherlands organization for scientific
research (NWO) VIDI grant (dossier 639.042.730).  A.~v.~E.\ was supported by the Beatrice and Vincent Tremaine Fellowship.

\appendix
\section{Flat-sky estimator}
\label{app:flatsky}

\subsection{Expressions}
\label{app:flatsky-expressions}

In this appendix, we present expressions corresponding to the estimator from Sec.~\ref{sec:estimator}, but in the flat-sky approximation. In this picture, the lensed and unlensed temperature fields are related by
\beq
\widetilde{T}(\vtheta) = T\!\lp \vtheta + \frac{\d\phi(\vtheta)}{\d\vtheta} \rp\ ,
\eeq
which can be written in Fourier space as
\beq
\widetilde{T}(\vl) \approx T(\vl) - \int_{\vl'} \vl'\cdot\lp \vl-\vl' \rp T(\vl') \phi(\vl-\vl')
\label{eq:ttildeflatsky}
\eeq
to first order in $\phi$, where $\vl$ is the Fourier conjugate of the 2d position-space coordinate $\vtheta$. A quadratic estimator for long modes of $T$ can be written as \cite{Hu:2001tn}
\beq
\widehat{T}(\vL) = \int_{\vl} g(\vl,\vL-\vl) T^{\rm obs}(\vl) \phi^{\rm obs}(\vL-\vl)\ ;
\label{eq:testflatsky}
\eeq
demanding that the estimator be unbiased and of minimum variance if the unlensed field is Gaussian fixes the the filter $g$ to be given by
\begin{align*}
&g(\vl,\vL-\vl) \\
&\quad \equiv N_L^{\widehat{T}\widehat{T}} 
	\frac{\vL\cdot\lp \vL-\vl \rp C_{|\vL-\vl|}^{\phi\phi}}
	{\lp C_{\ell}^{TT,\mathrm{res}} + N_{\ell}^{TT} \rp 
	\lp C_{|\vL-\vl|}^{\phi\phi}+N_{|\vL-\vl|}^{\phi\phi} \rp}\ ,
	\numberthis
\end{align*}
with reconstruction noise given by
\beq
N_L^{\widehat{T}\widehat{T}} \equiv \lb \int_{\vl}
	\frac{\lb \vL\cdot\lp \vL-\vl \rp C_{|\vL-\vl|}^{\phi\phi} \rb^2}
	{\lp C_{\ell}^{TT,\mathrm{res}} + N_{\ell}^{TT} \rp 
	\lp C_{|\vL-\vl|}^{\phi\phi}+N_{|\vL-\vl|}^{\phi\phi} \rp}
	 \rb^{-1}\ .
\eeq

\subsection{Reconstruction noise in large-$\ell$ limit}
\label{app:flatsky-noise}

In the typical case, the maximum wavenumber $\lmax$ used for the reconstruction will be much larger than the wavenumber $L$ of the mode being reconstructed, so it is instructive to examine the reconstruction noise in that limit~\cite{Bucher:2010iv,Hanson:2010rp}. Assuming $\lmax\gg L$, we find
\begin{align*}
N_L^{\widehat{T}\widehat{T}} &\approx 
	\lb \int_{\vl}  \lp \vL\cdot\vl \rp^2 
	\frac{C_\ell^{\phi\phi}}{C_{\ell}^{TT,\mathrm{res}} + N_{\ell}^{TT}}
	\frac{C_\ell^{\phi\phi}}{C_{\ell}^{\phi\phi} + N_{\ell}^{\phi\phi}} \rb^{-1} \\
&= \lb \frac{L^2}{4\pi} \int d\ell\,\ell^3 
	\frac{C_\ell^{\phi\phi}}{C_{\ell}^{TT,\mathrm{res}} + N_{\ell}^{TT}}
	\frac{C_\ell^{\phi\phi}}{C_{\ell}^{\phi\phi} + N_{\ell}^{\phi\phi}} \rb^{-1}\ ,
	\numberthis
	\label{eq:recnoiseflatsky}
\end{align*}
revealing that the noise on $L^2 C_L^{\widehat{T}\widehat{T}}$ is white in this limit. Furthermore, this expression shows that scales for which~$\phi$ is noise-dominated can still contribute significantly to the reconstruction of long temperature modes. This is due to the steep scaling of the ratio $C_\ell^{\phi\phi}/C_\ell^{TT}$ at small scales: in the absence of lensing or kSZ, $C_\ell^{\phi\phi}/C_\ell^{TT} \sim \ell^6$ for $3000\lesssim\ell\lesssim 6000$, while for the lensed temperature spectrum with kSZ we find $C_\ell^{\phi\phi}/C_\ell^{TT} \sim \ell^{1.5}$ for $2000\lesssim\ell\lesssim 3500$. Thus, on scales where~$\phi$ is noise-dominated but $T$ is signal-dominated, the integral in Eq.~\eqref{eq:recnoiseflatsky} can still have a significant contribution if the $\phi$ noise is not exponentially increasing on those scales.

This is the case for CMB-S4:~$\phi$ is noise-dominated for $\ell\gtrsim 1000$, but the noise increases rather slowly at smaller scales ($N_{\ell}^{\phi\phi} \sim \ell^{0.8}$ for $1000\lesssim\ell\lesssim 4000$), so the signal to noise on $\widehat{T}$ continues to increase with $\ell$ until $\ell\approx 3500$, when~$T$ becomes noise-dominated. Numerical computations reveal that the $T$ reconstruction noise for CMB-S4 is roughly equivalent to that for an experiment that can measure $T$ and $\phi$ to cosmic variance limits up to $\lmax\approx 2000$.

We can contrast this situation with  the flat-sky reconstruction noise for the standard quadratic lensing estimator~\cite{Hu:2001tn}:
\beq
N_L^{\widehat{\phi}\widehat{\phi}} \approx \lb \frac{L^4}{4\pi} \int d\ell\,\ell 
	\lp \frac{C_\ell^{TT}}{C_\ell^{TT}+N_\ell^{TT}} \rp^2 \rb^{-1}\ .
	\label{eq:phirecnoise}
\eeq
Since $N_\ell^{TT}/C_\ell^{TT}$ typically increases exponentially for scales smaller than the beam scale, this shows that wavenumbers for which $N_\ell^{TT} \gtrsim C_\ell^{TT}$ will contribute negligibly to the lensing reconstruction.

\begin{widetext}
\section{Contribution of secondary CMB effects to estimator}
\label{app:secondary}

The estimator in Sec.~\ref{sec:estimator} assumed that the entire temperature field is lensed by the same lensing potential, but secondary effects generated at lower redshifts will be lensed differently, leading to biases on the reconstructed modes. In this appendix, we derive expressions for these biases, in both the full-sky and flat-sky formalisms.

\subsection{Full-sky expressions}

In the presence of a secondary effect $X$ that is sourced over redshifts between the observer and the last-scattering surface, Eq.~\eqref{eq:tonereconstruction} for the lensed temperature field $\widehat{T}$ will be modified as
\beq
\widetilde{T}_{\ell_1 m_1}^* = \sum_{\ell_2 m_2 \ell_3 m_3} 
	\Gamma^{\ell_1 \ell_2 \ell_3}_{m_1 m_2 m_3} 
	\lb \phi_{\ell_2 m_2}(z_*) T_{\ell_3 m_3}^{\rm p}
	+ \int_0^{z_*} dz\, \phi_{\ell_2 m_2}(z) \frac{\d T_{\ell_3 m_3}^X}{\d z} \rb \ ;
	\numberthis
\eeq
that is, the contribution to $T_{\ell m}^X$ from a redshift interval $dz$ centered at $z$ will be lensed by the lensing potential out to~$z$. This will contribute an additive bias to the expectation value of the estimator in Eq.~\eqref{eq:test}, given by
\beq
\Delta\left\la \widehat{T}_{LM} \right\ra = 
	N_L^{\widehat{T} \widehat{T}} \sum_{\ell_1\ell_2} S_{\ell_1\ell_2 L}
	\int_0^{z_*} C_{\ell_2}^{\phi\phi}(z,z_*) \frac{\d T_{LM}^X}{\d z}\ ,
	\label{eq:testbias}
\eeq
and this will in turn bias the power spectrum of the estimator:
\beq
\Delta C_L^{\widehat{T}\widehat{T}}
	= \sum_{\ell_1\ell_2} S_{\ell_1\ell_2 L} \sum_{\ell_3\ell_4}  S_{\ell_3\ell_4 L} \\
	\int_0^{z_*} dz\, C_{\ell_2}^{\phi\phi}(z,z_*)
	\int_0^{z_*} dz' C_{\ell_4}^{\phi\phi}(z',z_*) C_L^{\d X\d X}(z,z')\ .
	\label{eq:clttbias}
\eeq
In Eqs.~\eqref{eq:testbias} and~\eqref{eq:clttbias}, we have defined $S_{\ell_1\ell_2 L}$ as
\beq
S_{\ell_1\ell_2 L}
	\equiv   e_{\ell_1 \ell_2 L} 
	\frac{(2\ell_1+1)(2\ell_2+1)}{4\pi} 
	\lp J_{\ell_1 \ell_2 L}\rp^2 
 	\lp \frac{1}{C_{\ell_1}^{TT,\mathrm{res}} + N_{\ell_1}^{TT}}\rp
	 \lp \frac{C_{\ell_2}^{\phi\phi} }{C_{\ell_2}^{\phi\phi}+N_{\ell_2}^{\phi\phi}}\rp .
	 \numberthis
\eeq
(With this definition, $N_L^{\widehat{T}\widehat{T}}$ from Eq.~\eqref{eq:reconstructionnoise} becomes $\lb \sum_{\ell_1 \ell_2} S_{\ell_1\ell_2 L} C_{\ell_2}^{\phi\phi} \rb^{-1}$.) Also, we have written $C_{\ell}^{\phi\phi}(z,z_*)$ for the cross power spectrum between lensing potentials for sources at $z$ and $z_*$,
\beq
C_{\ell}^{\phi\phi}(z,z_*) \equiv \left\la \phi_{\ell m}(z) \phi^*_{\ell',m'}(z_*) \right\ra\ ,
\eeq
and $C_L^{\d X\d X}(z,z')$ for the cross power spectrum between contributions to $T^X$ from redshifts $z$ and $z'$:
\beq
C_L^{\d X\d X}(z,z') \equiv \left\la \frac{\d T_{LM}^X}{\d z} \frac{\d T_{LM}^{X*}}{\d z'} \right\ra\ .
\eeq
If $T^X$ can be written as a projection of the matter overdensity $\delta$ against a window function $W^X$,
\beq
T^X(\nhat) = \int_0^{z_*} dz\, W^X(\d^2,z)\delta(\chi[z]\nhat;z)\ ,
\eeq
then $C_L^{\d X\d X}(z,z')$ can be written as
\beq
C_L^{\d X\d X}(z,z') = \frac{2}{\pi} \int dk\, k^2 
	W^X(-k^2,z) W^X(-k^2,z') j_L(k\chi[z]) j_L(k\chi[z']) D(z) D(z') P_{\rm lin}(k;z=0)\ ,
	\label{eq:CLdxdx}
\eeq
assuming we work on scales where linear theory is a good description of the matter field. In the Limber approximation~\cite{Limber:1953,LoVerde:2008re}, Eq.~\eqref{eq:CLdxdx} simplifies to
\beq
C_L^{\d X\d X}(z,z') \approx \delta(z-z') \frac{H(z)}{\chi[z]^2}
	\lb W^X(-k^2,z)D(z) \rb^2 \left. P_{\rm lin}\lp k;z=0\rp \right|_{k=(L+1/2)/\chi[z]}\ .
	\label{eq:CLdxdxLimber}
\eeq
For the ISW effect, the window function is
\beq
W^{\rm ISW}(-k^2,z) = \frac{3\Omm H_0^2}{c^2k^2} \lb 1-f(z) \rb\ .
\eeq

\subsection{Flat-sky expressions}

In the flat-sky approximation, the lensed temperature in Eq.~\eqref{eq:ttildeflatsky} is modified to
\beq
\widetilde{T}(\vl) \approx T(\vl) - \int_{\vl'} \vl'\cdot\lp \vl-\vl' \rp 
	\lb \phi(\vl-\vl';z_*) T^{\rm p}(\vl')  + \int_0^{z_*} dz\, \phi(\vl-\vl';z) \frac{\d T^X(\vl')}{\d z}\rb\ ,
\eeq
resulting in a bias on the estimator in Eq.~\eqref{eq:testflatsky} of the form
\beq
\Delta\left\la \widehat{T}(\vL) \right\ra = 
	\int_{\vl} g(\vl,\vL-\vl) \vL\cdot(\vL-\vl) 
	\int_0^{z_*} C_{|\vL-\vl|}^{\phi\phi}(z,z_*) \frac{\d T^X(\vL)}{\d z}\ .
\eeq
The corresponding bias on the reconstructed power spectrum is
\begin{align*}
\Delta C_L^{\widehat{T}\widehat{T}}
	&= \int_{\vl} g(\vl,\vL-\vl) \vL\cdot(\vL-\vl) \int_{\vl'} g(\vl',\vL-\vl') \vL\cdot(\vL-\vl') \\
&\quad\times\int_0^{z_*} dz\, C_{|\vL-\vl|}^{\phi\phi}(z,z_*)
	\int_0^{z_*} dz' C_{|\vL-\vl'|}^{\phi\phi}(z',z_*) C_L^{\d X\d X}(z,z')\ .
	\numberthis
\end{align*}
Using the Limber expression from Eq.~\eqref{eq:CLdxdxLimber} and taking the $\ell,\ell'\gg L$ limit allows for an approximation for this bias to be efficiently computed numerically.

\end{widetext}

\bibliography{references}
\end{document}